\documentclass[a4paper]{article}

\usepackage{INTERSPEECH2021}
\usepackage{subfigure}

\usepackage{amssymb}
\usepackage{multirow}
\usepackage{graphicx}
\usepackage{tabularx}
\usepackage{hyperref}
\usepackage[ruled,vlined]{algorithm2e}
\usepackage[table,xcdraw]{xcolor}

\newlength\mylength
\title{On the Design of Deep Priors for Unsupervised Audio Restoration}
\name{Vivek Sivaraman Narayanaswamy$^1$, Jayaraman J. Thiagarajan$^2$, Andreas Spanias$^1$}
\address{
  $^1$Arizona State University\\
  $^2$Lawrence Livermore National Laboratory}
\email{vnaray29@asu.edu, jjayaram@llnl.gov, spanias@asu.edu}

\begin{document}

\maketitle
\begin{abstract}
  Unsupervised deep learning methods for solving audio restoration problems extensively rely on carefully tailored neural architectures that carry strong inductive biases for defining priors in the time or spectral domain. In this context, lot of recent success has been achieved with sophisticated convolutional network constructions that recover audio signals in the spectral domain. However, in practice, audio priors require careful engineering of the convolutional kernels to be effective at solving ill-posed restoration tasks, while also being easy to train. To this end, in this paper, we propose a new U-Net based prior that does not impact either the network complexity or convergence behavior of existing convolutional architectures, yet leads to significantly improved restoration. In particular, we advocate the use of carefully designed dilation schedules and dense connections in the U-Net architecture to obtain powerful audio priors.  Using empirical studies on standard benchmarks and a variety of ill-posed restoration tasks, such as audio denoising, in-painting and source separation, we demonstrate that our proposed approach consistently outperforms widely adopted audio prior architectures.


\end{abstract}

\noindent\textbf{Index Terms}: Deep audio priors, unsupervised audio restoration, denoising, inpainting, source separation.

\section{Introduction}
Deep convolutional neural networks have proven to be effective for recovering signals from noisy observations. Consequently, state-of-the-art solutions for challenging problems such as enhancement~\cite{segan}, inpainting~\cite{chang2019deep} and source separation~\cite{luo2019conv} are based on convolutional architectures~\cite{giri2019attention, defossez2019demucs}. While majority of this success has been with supervisory data, recent focus has shifted to unsupervised approaches that do not require expensive data collection and curation. Given the ill-posed nature of audio restoration, choice of suitable audio priors is critical to the success of unsupervised learning approaches. 

The seminal work of Ulyanov \textit{et al.} introduced the notion of \textit{deep image priors} and showed that convolutional neural network architectures can provide powerful signal priors for solving image restoration problems. In contrast to unsupervised approaches that use data-driven priors based on pre-trained generative models (e.g, Generative Adversarial Networks (GANs))~\cite{shah2018solving, narayanaswamy2020unsupervised} to solve inverse problems, these structural priors do not require any training data and the optimization can be carried out using a single observation. The flexibility and the effectiveness of this approach has motivated the design of suitable priors for audio restoration tasks. A number of recent studies~\cite{zhang2019deep, tian2019deep, michelashvili2019speech,chang2019deep} have showed that different variants of convolutional architectures are highly effective choices. For example, Michelashvili et.al~\cite{michelashvili2019speech} used the Wave-U-Net~\cite{stoller2018wave} architecture to denoise audio signals. Interestingly,  convolutional network constructions that operate in the spectral domain have been found to be consistently superior. For example, Tian \textit{et al.}~\cite{tian2019deep} proposed Deep Audio Priors, that utilize separate randomly initialized U-Net models~\cite{ronneberger2015u} to obtain time-frequency masks and audio source estimates respectively for source separation without any pre-training. 

Deep audio priors can be characterized using a number of factors including recovery performance across different inversion tasks, ease of training, and computational efficiency. For example, replacing standard convolutions with dilated convolutions is known to improve recovery performance without any impact on the computational efficiency. More recently, Zhang \textit{et al.}~\cite{zhang2019deep} explored the use of harmonic convolutions that carefully engineer the convolutional kernels to better capture multi-scale harmonic structure in audio. Takeuchi \textit{et al.}~\cite{takeuchi2020harmonic} subsequently improved the computational efficiency of harmonic convolutions through harmonic lowering. However, audio priors based on harmonic convolutions require significantly larger number of training iterations compared to standard convolutions. While this challenge was addressed in~\cite{zhang2019deep} through the use of multiple anchors for harmonic convolutions, the resulting audio prior can be of significantly higher complexity.

In this paper, we revisit the design of audio priors (spectral domain) and propose a new U-Net based prior, that does not impact either the network complexity or the convergence behavior, but consistently leads to high-fidelity restoration. We find that unsupervised audio restoration can be improved by adopting dilated convolutions with an exponentially increasing dilation schedule and by introducing dense connections. Using empirical studies on audio denoising, inpainting and source separation experiments, we show that the proposed audio prior\footnote{The codes for our implementation are publicly available: \url{https://github.com/vivsivaraman/designaudiopriors}} better extracts multi-scale features from time-frequency representations of audio signals, significantly outperforms widely adopted deep audio priors, and is computationally efficient when compared to harmonic convolutions~\cite{zhang2019deep, takeuchi2020harmonic}.

\begin{figure*}
    \centering
    \includegraphics[width=0.99\textwidth]{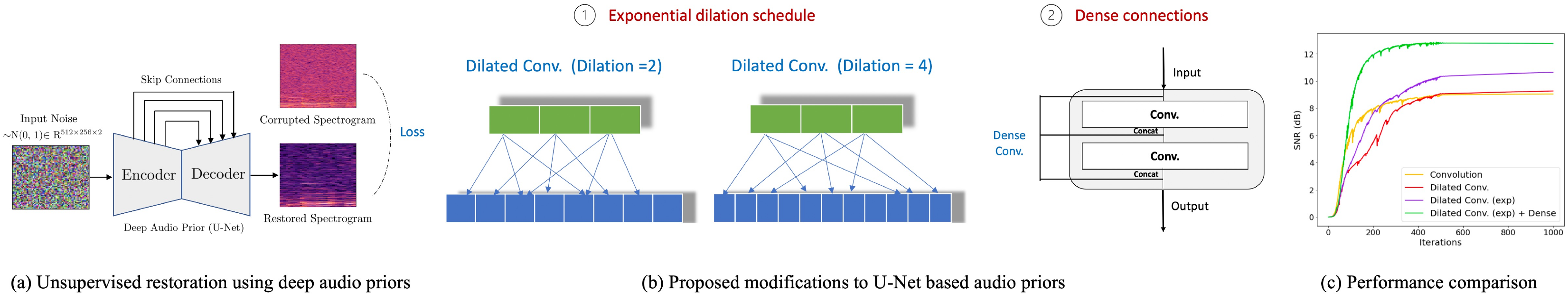}
    \caption{We propose a new deep audio prior construction that is well suited for challenging unsupervised restoration tasks. Through the use of dilated convolutions with a carefully engineered dilation schedule and dense connections in a standard U-Net, we achieve significant performance gains over state-of-the-art approaches. The example in (c) corresponds to an audio denoising experiment.}
    \label{fig:teaser}
\end{figure*}


\section{Unsupervised Audio Restoration}
Audio restoration refers to the process of recovering an audio signal $\mathbf{x} \in \mathbb{R}^{m \times c}$ from a corrupted observation $\hat{\mathbf{x}}\in \mathbb{R}^{n \times c}$~\cite{spanias2006audio}. Here $m$ and $n$ denote the length of the observations while $c$ denotes the number of channels. Without loss of generality, in this work, we assume $m = n$ and the signals to be mono-channel i.e., $c = 1$. In this paper, we consider three popular audio restoration tasks, namely audio denoising, inpainting and source separation. Audio denoising refers to the task of removing noise from a corrupted audio while preserving the underlying characteristics. On the other hand, audio inpainting attempts to recover the original signal from observations that are spatio-temporally masked, and is typically utilized for audio editing and packet loss recovery in receiver systems. Finally, source separation refers to the process of recovering the constituent audio sources present in a given mixture observation, wherein the mixing process may not be known in advance. 

In practice, since the corruption process (e.g., type of noise or noise level) is unknown \textit{a} priori, audio restoration is a severely ill-posed inverse problem and often requires meaningful signal priors~\cite{ulyanov2018deep,zhang2019deep}. In this context, deep audio priors have become highly prevalent, particularly for unsupervised restoration. Formally, given a corrupted observation $\hat{\mathbf{x}}$ and an untrained convolutional neural network $f_\Theta$ with parameters $\Theta$, the structure of the neural network can innately regularize the inverse optimization. The intuition behind such structural priors is that if the network is capable of modeling the signal priors induced by its structure, the network would fit the signal easily than that noise. In a deep audio prior-based restoration, the clean signal can be directly obtained as $f_{\Theta}(\mathbf{z})$, where $\mathbf{z}$ is a random noise (latent) vector drawn from a known distribution. 

\noindent\textbf{Proposed Work.} In this work, we study the design of effective deep audio priors for practical restoration tasks. Though existing efforts in the literature have explored the use of dilated and harmonic convolutions in U-Net based priors, large performance gains and desirable training behavior were enabled only by increasing the complexity of the prior, e.g., multiple anchors for harmonic convolutions~\cite{zhang2019deep}. In contrast, we propose key modifications to U-Net based audio priors that do not significantly increase the network complexity, but can produce large performance gains in restoration tasks. More specifically, we propose the use of exponentially increasing receptive fields via dilated convolutions~\cite{oord2016wavenet} by adopting a pre-specified dilation schedule which dispenses the need for explicit resampling techniques for better feature extraction. Furthermore, we introduce dense connections between within each layer, as well as between \textit{upstream} and \textit{downstream} paths of the U-Net to promote better feature reuse and improved gradient flows. Together, deep audio priors with these two modifications consistently outperform other widely adopted prior choices.

\section{Proposed Approach}

Figure \ref{fig:teaser} provides an overview of our approach. We propose a new U-Net based deep audio prior construction that we empirically find to be superior to existing convolutional architectures for unsupervised restoration. In this section, we describe the key steps  of our approach: (i) designing an U-Net architecture; (ii) using dilated convolutions with a specific dilation schedule; and (iii) adding dense connections for improved gradient flow.

\subsection{U-Net Architecture Design}
We adopt the U-Net architecture as a structural prior to effectively regularize the ill-posed tasks of audio restoration. The architecture is comprised of two convolutional blocks in the \textit{downstream} path where each block in turn contains two $2$D convolution layers with filter sizes $\{35, 70\}$ and $\{70, 140\}$ respectively. Correspondingly, the \textit{upstream} path is comprised of two convolutional blocks, wherein each block contains a bi-linear upsampling step followed by two convolution layers with filter sizes $\{140, 70\}$ and $\{70, 35\}$ respectively. The bottleneck block between \textit{downstream} and \textit{upstream} paths consists of two more convolutional layers with $70$ filters each. The final output is obtained using another convolutional layer with the desired number of channels. In addition, skip connections are included between the convolutional blocks in the downstream and upstream paths, which combine the coarse and fine grained features from the respective paths to improve signal reconstruction.  

\subsection{Dilated Convolutions with an Exponential Schedule}
The success of the audio prior relies heavily on the quality of the features extracted at different scales for signal reconstruction. Recovering audio signals can be challenging due to the inherent periodicities and complex spatio-temporal statistics, and this naturally motivates feature extraction strategies that can leverage information over wider receptive fields at increasing depths. To this end, we introduce dilations in all convolutional layers of the U-Net, wherein the dilation rates are exponentially increased for each subsequent convolution layer (in factors of 2). Specifically, starting with a dilation factor of $2$ for the first convolution layer in the first block, the dilation rate grows upto $32$ in the bottleneck block. The \textit{upstream} is correspondingly designed to mirror the \textit{downstream} architecture. The inherent downsampling operation in the \textit{downstream} path (max-pooling) combined with the exponential dilation schedule effectively enables feature extraction across significantly large receptive fields (e.g., periodicities).

\subsection{Adding Dense Connections}
In addition to enabling multi-scale feature extraction via an exponential dilation schedule, we aim to enhance the U-Net architecture further by adding dense connections in order to encourage feature reuse and improve gradient flow even at increasing layer depths (see Fig.\ref{fig:teaser}(b)). More specifically, we include dense connections between convolutional layers within each convolutional block, i.e., the feature maps produced by each layer are concatenated to the subsequent layers in the block. In order to prevent the accumulation of a large number of feature maps at increasing depths, following Thiagarajan \textit{et al.}~\cite{thiagarajan2020ddxnet}, we include a transition block (implemented using a single standard $2-$D convolutional layer) at the end of every dense block, which reduces the dimensionality of the resulting feature maps. 

\begin{figure}
    \centering
    \includegraphics[width=0.8\columnwidth]{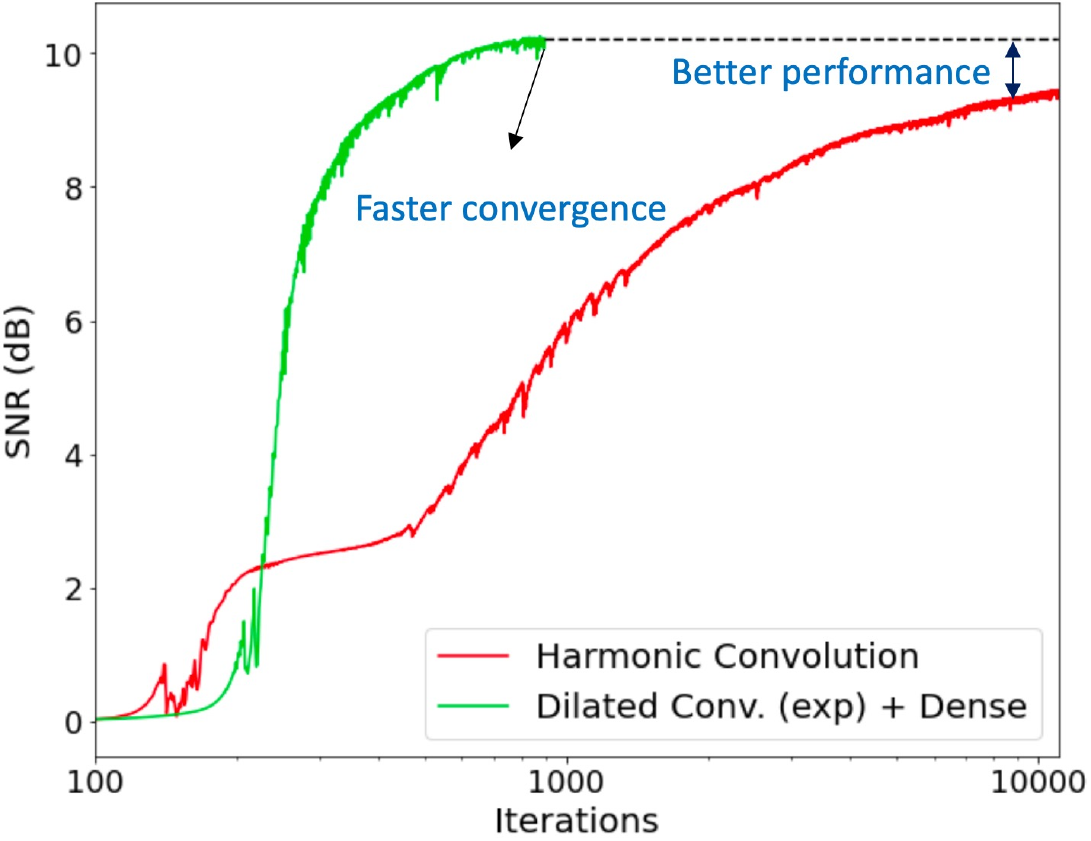}
    \caption{Comparing the convergence of the proposed audio prior to U-Nets based on harmonic convolutions. Our prior achieves both significantly faster convergence and marginal performance gains over the latter, while convincingly outperforming other widely adopted deep audio prior constructions.}
    \label{fig:comp}
\end{figure}

\noindent \textbf{Comparison to Harmonic Convolutions.} Recent efforts~\cite{zhang2019deep} have recommended harmonic convolutions as an effective choice over standard convolutions for designing audio priors. However, as illustrated in Figure \ref{fig:comp} for an audio denoising example, the proposed audio prior construction requires significantly lower number of iterations ($10\times$) to converge when compared to harmonic convolutions, while also providing non-trivial gains in the restoration performance. 

\section{Experiments}
\begin{table*}[!t]
\centering
\renewcommand*{\arraystretch}{1.2}
	\caption{Audio denoising performance of deep audio priors under the presence of Gaussian noise.}
	\resizebox{0.8\textwidth}{!}{%
\begin{tabular}{|l|c|c|c|c|c|}
\hline
\rowcolor[HTML]{C0C0C0} 
\multicolumn{1}{|c|}{\cellcolor[HTML]{C0C0C0}}                                      & \multicolumn{2}{c|}{\cellcolor[HTML]{C0C0C0}\textbf{LJ-Speech}}                                                         & \multicolumn{2}{c|}{\cellcolor[HTML]{C0C0C0}\textbf{Digits}}                                                            & \multicolumn{1}{c|}{\cellcolor[HTML]{C0C0C0}\textbf{Piano}} \\ \cline{2-6} 
\rowcolor[HTML]{EFEFEF} 
\multicolumn{1}{|c|}{\multirow{-2}{*}{\cellcolor[HTML]{C0C0C0}\textbf{DAP Design}}} & \multicolumn{1}{c|}{\cellcolor[HTML]{EFEFEF}\textbf{PESQ}} & \multicolumn{1}{c|}{\cellcolor[HTML]{EFEFEF}\textbf{PSNR}} & \multicolumn{1}{c|}{\cellcolor[HTML]{EFEFEF}\textbf{PESQ}} & \multicolumn{1}{c|}{\cellcolor[HTML]{EFEFEF}\textbf{PSNR}} & \multicolumn{1}{c|}{\cellcolor[HTML]{EFEFEF}\textbf{PSNR}}  \\ \hline \hline
Convolution                                                                         & 1.73 $\pm$ 0.17 & 6.85 $\pm$ 1.47 & 1.99 $\pm$ 0.47 & 9.54 $\pm$ 2.96 & 9.17 $\pm$ 1.34                                             \\ \hline
Dilated Conv.                                                                        & 1.76 $\pm$ 0.21 & 7.18 $\pm$ 1.55 & 2.08 $\pm$ 0.47 & 10.85 $\pm$ 2.72 & 9.11 $\pm$ 1.49                                             \\ \hline
Dilated Conv. (exp)                                                                 & 2.00 $\pm$ 0.19 & 7.68 $\pm$ 1.76 & 2.20 $\pm$ 0.45 & 11.52 $\pm$ 2.89 & 10.68 $\pm$ 1.47                                             \\ \hline
Dilated Conv. (exp) + Dense                                                         & \textbf{2.07 $\pm$ 0.16} & \textbf{8.15 $\pm$ 1.94} & \textbf{2.23 $\pm$ 0.47} & \textbf{11.55 $\pm$ 2.91} & \textbf{12.50 $\pm$ 1.11}                                    \\ \hline
\end{tabular}
}
\label{table:den_g}
\end{table*}

\begin{table*}[!t]
\centering
\renewcommand*{\arraystretch}{1.2}
	\caption{Audio denoising performance of deep audio priors under the presence of environmental noise.}
	\resizebox{0.8\textwidth}{!}{%
\begin{tabular}{|l|c|c|c|c|c|}
\hline
\rowcolor[HTML]{C0C0C0} 
\multicolumn{1}{|c|}{\cellcolor[HTML]{C0C0C0}}                                      & \multicolumn{2}{c|}{\cellcolor[HTML]{C0C0C0}\textbf{LJ-Speech}}                                                         & \multicolumn{2}{c|}{\cellcolor[HTML]{C0C0C0}\textbf{Digits}}                                                            & \multicolumn{1}{c|}{\cellcolor[HTML]{C0C0C0}\textbf{Piano}} \\ \cline{2-6} 
\rowcolor[HTML]{EFEFEF} 
\multicolumn{1}{|c|}{\multirow{-2}{*}{\cellcolor[HTML]{C0C0C0}\textbf{DAP Design}}} & \multicolumn{1}{c|}{\cellcolor[HTML]{EFEFEF}\textbf{PESQ}} & \multicolumn{1}{c|}{\cellcolor[HTML]{EFEFEF}\textbf{PSNR}} & \multicolumn{1}{c|}{\cellcolor[HTML]{EFEFEF}\textbf{PESQ}} & \multicolumn{1}{c|}{\cellcolor[HTML]{EFEFEF}\textbf{PSNR}} & \multicolumn{1}{c|}{\cellcolor[HTML]{EFEFEF}\textbf{PSNR}}  \\ \hline \hline
Convolution                                                                          & 1.91 $\pm$ 0.26 & 4.36 $\pm$ 1.29 & 2.23 $\pm$ 0.58 & 6.39 $\pm$ 1.81 & 5.73 $\pm$ 1.06                                             \\ \hline
Dilated Conv. (exp)                                                                 & 2.04 $\pm$ 0.24 & 5.03 $\pm$ 1.27 & 2.31 $\pm$ 0.69 & 6.95 $\pm$ 1.92 & 6.40 $\pm$ 1.12                                             \\ \hline
Dilated Conv. (exp) + Dense                                                         & \textbf{2.31 $\pm$ 0.22} & \textbf{5.58 $\pm$ 1.34} & \textbf{2.46 $\pm$ 0.58} & \textbf{7.19 $\pm$ 1.82} & \textbf{7.30 $\pm$ 1.12}                                    \\ \hline
\end{tabular}
}
\label{table:den_e}
\end{table*}

\begin{table*}[!t]
\centering
\renewcommand*{\arraystretch}{1.2}
	\caption{Audio inpainting performance of deep audio priors under random spatio-temporal masking.}
	\resizebox{0.7\linewidth}{!}{%
\begin{tabular}{|l|c|c|c|c|}
\hline
\rowcolor[HTML]{C0C0C0} 
\multicolumn{1}{|c|}{\cellcolor[HTML]{C0C0C0}}                                      & \multicolumn{2}{c|}{\cellcolor[HTML]{C0C0C0}\textbf{LJ-Speech}}                                                         & \multicolumn{2}{c|}{\cellcolor[HTML]{C0C0C0}\textbf{Digits}}                                                            \\ \cline{2-5} 
\rowcolor[HTML]{EFEFEF} 
\multicolumn{1}{|c|}{\multirow{-2}{*}{\cellcolor[HTML]{C0C0C0}\textbf{DAP Design}}} & \multicolumn{1}{c|}{\cellcolor[HTML]{EFEFEF}\textbf{Spec. SNR}} & \multicolumn{1}{c|}{\cellcolor[HTML]{EFEFEF}\textbf{Env. Dist.}} & \multicolumn{1}{c|}{\cellcolor[HTML]{EFEFEF}\textbf{Spec. SNR}} & \multicolumn{1}{c|}{\cellcolor[HTML]{EFEFEF}\textbf{Env. Dist.}} \\ \hline \hline
{Convolution}    & 7.27 $\pm$ 1.43 & 0.08 $\pm$ 0.02 & 7.34 $\pm$ 1.91 & 0.12 $\pm$ 0.03                                    \\ \hline
{Dilated Conv.} & 7.18 $\pm$ 1.06 & 0.08 $\pm$ 0.02 & 7.78 $\pm$ 1.88 & 0.11 $\pm$ 0.03                                             \\ \hline
{Dilated Conv. (exp)} & 7.95 $\pm$ 1.16 & 0.07 $\pm$ 0.02  & 9.02 $\pm$ 1.96 & 0.10 $\pm$ 0.03                                             \\ \hline
{Dilated Conv. (exp) + Dense}& \textbf{10.01 $\pm$ 1.78} & \textbf{0.06 $\pm$ 0.02}  & \textbf{10.80 $\pm$ 2.52}  & \textbf{0.09 $\pm$ 0.03}                                \\ \hline
\end{tabular}
}
\label{table:inp}
\end{table*}

\begin{table*}[!t]
\centering
\renewcommand*{\arraystretch}{1.5}
	\caption{Unsupervised source separation performance of deep audio priors.}
	\resizebox{\textwidth}{!}{%
\begin{tabular}{|l|l|l|l|l|l|l|l|l|}
\hline
\rowcolor[HTML]{C0C0C0} 
\multicolumn{1}{|c|}{\cellcolor[HTML]{C0C0C0}{\color[HTML]{FFFFFF} }}                                  & \multicolumn{2}{c|}{\cellcolor[HTML]{C0C0C0}\textbf{SDR (dB)}}                                                            & \multicolumn{2}{c|}{\cellcolor[HTML]{C0C0C0}\textbf{SIR (dB)}}                                                            & \multicolumn{2}{c|}{\cellcolor[HTML]{C0C0C0}\textbf{Spec. SNR (dB)}}                                                      & \multicolumn{2}{c|}{\cellcolor[HTML]{C0C0C0}\textbf{Env. Dist}}                                                           \\ \cline{2-9} 
\rowcolor[HTML]{C0C0C0} 
\multicolumn{1}{|c|}{\multirow{-2}{*}{\cellcolor[HTML]{C0C0C0}{ \textbf{DAP Design}}}} & \multicolumn{1}{c|}{\cellcolor[HTML]{EFEFEF}\textbf{Piano}} & \multicolumn{1}{c|}{\cellcolor[HTML]{EFEFEF}\textbf{Drums}} & \multicolumn{1}{c|}{\cellcolor[HTML]{EFEFEF}\textbf{Piano}} & \multicolumn{1}{c|}{\cellcolor[HTML]{EFEFEF}\textbf{Drums}} & \multicolumn{1}{c|}{\cellcolor[HTML]{EFEFEF}\textbf{Piano}} & \multicolumn{1}{c|}{\cellcolor[HTML]{EFEFEF}\textbf{Drums}} & \multicolumn{1}{c|}{\cellcolor[HTML]{EFEFEF}\textbf{Piano}} & \multicolumn{1}{c|}{\cellcolor[HTML]{EFEFEF}\textbf{Drums}} \\ \hline \hline
Convolution                                                                                            & 2.56 $\pm$ 1.71                                             & -0.28 $\pm$ 1.45                                            & 13.17 $\pm$ 6.62                                            & -6.41 $\pm$8.18                                             & 2.75 $\pm$ 1.59                                             & 0.01 $\pm$ 0.97                                             & 0.28 $\pm$ 0.09                                             & 0.18 $\pm$ 0.07                                             \\ \hline
Dilated Conv.                                                                                          & 2.54 $\pm$ 1.63                                             & 0.02 $\pm$ 1.47                                             & 13.09 $\pm$ 7.47                                            & -3.41 $\pm$ 9.09                                            & 2.75 $\pm$ 1.39                                             & 0.04 $\pm$ 0.74                                             & 0.26 $\pm$ 0.08                                             & 0.15 $\pm$ 0.05                                             \\ \hline
Dilated Conv. (exp)                                                                                    & 3.07 $\pm$ 1.38                                             & 0.15 $\pm$ 1.87                                             & 11.84 $\pm$ 8.64                                            & 1.12 $\pm$ 5.25                                             & 3.26 $\pm$ 1.64                                             & 0.17 $\pm$ 1.65                                             & 0.25 $\pm$ 0.06                                             & \textbf{0.14 $\pm$ 0.05}                                    \\ \hline
Dilated Conv. (exp) + Dense                                                                            & \textbf{4.84 $\pm$ 2.61}                                    & \textbf{0.61 $\pm$ 3.09}                                    & \textbf{12.57 $\pm$ 7.62}                                   & \textbf{1.93 $\pm$ 5.64}                                    & \textbf{5.43 $\pm$ 2.04}                                    & \textbf{0.54 $\pm$ 1.77}                                    & \textbf{0.21 $\pm$ 0.08}                                    & \textbf{0.14 $\pm$ 0.04}                                    \\ \hline
\end{tabular}
}
\label{table:source}
\end{table*}

In this section, we present empirical studies to evaluate our proposed approach on three ill-posed audio restoration tasks, namely denoising, in-painting and source separation. We will begin by discussing the datasets used in our study. 

\noindent\textbf{Datasets.} We used the following datasets for our study: LJSpeech, SC09 Spoken Digit (SC09), drum and piano sounds. LJSpeech~\cite{ljspeech17} is an open source dataset containing audio clips of duration $\sim8s$ at 22kHz of a speaker reading sentences. The SC09 dataset~\cite{warden2018speech, donahue2018adversarial} is comprised of spoken digits (0-9) with duration $\sim 1s$ at 16kHz. The drum sounds dataset~\cite{donahue2018adversarial}
contains single drum hit audio of duration $\sim1s$ at 16kHz, while the piano dataset~\cite{donahue2018adversarial} contains clips of duration $>50s$ at 48kHz.

\noindent\textbf{Pre-processing.} In all our experiments, we resample the audio samples to 16kHz and use clips of duration $2s$ for LJSpeech and $1s$ for other datasets. We carry then compute the spectrograms for the audio clips, using window length $1022$ and hop length $64$. Following Zhang~\cite{zhang2019deep} \textit{et al.}, we utilize both the real and imaginary parts of the spectrogram as a $2-$channel input. 

\noindent\textbf{Baselines}. We compare the performance of our audio prior to the widely adopted U-Net priors based on regular convolutions and dilated convolutions (constant dilation factor). For ablation, we also considered a variation where we used the exponential dilation schedule without dense connections. Though harmonic convolution~\cite{zhang2019deep} is another choice for implementing the audio prior, due to its significantly slower convergence (see Figure \ref{fig:comp}), we did not include it as a baseline approach. However, from our experiments, we found that our proposed approach consistently outperformed U-Nets with harmonic convolutions. 

\subsection{Audio Denoising}
In this task, using single corrupted observation $\hat{\mathbf{x}}$, we use deep audio priors to recover the underlying clean signal ${\mathbf{x}}$: $$\min_{\Theta} \mathcal{L}(f_\Theta(\mathbf{z}), \hat{\mathbf{x}}),$$ where $f_\Theta(\mathbf{z})$ is the restored output from the audio prior parameterized by $\Theta$, and $\mathcal{L}$ is implemented as the $\ell_2$ loss. We evaluate the proposed audio prior under two different noise scenarios (i)~\textit{Gaussian Noise:} We add Gaussian noise with standard deviation 0.1 to clean audio; (ii)~\textit{Environmental Noise:} We used Living Room and Traffic Noise samples from the DEMAND database~\cite{thiemann2013diverse} and synthesize observations by adding them with the clean audio  at SNRs chosen randomly between $5$ and $9$dB. We performed the optimization on each observation for $2000$ iterations using the ADAM optimizer and learning rate $0.001$. 

\noindent\textbf{Metrics}. Follow standard practice, we used the PESQ (Perceptual Evaluation of Speech Quality) and the PSNR (Peak-Signal to Noise Ratio) metrics.

\noindent\textbf{Findings}. Tables \ref{table:den_g} and \ref{table:den_e} show the performance of our approach against the baseline audio prior constructions on both noise settings. We report the performance metrics obtained on 50 random samples from each of the datasets. We find that our proposed approach provides a significantly superior performance over standard convolutions even under challenging environmental noise conditions. Note that, while dilated convolutions with a constant dilation factor are better than regular convolutions, the exponential dilation schedule improves by a bigger margin.

\subsection{Audio In-painting}
In this task, we use deep audio priors to fill masked regions in the observation $\hat{\mathbf{x}}$ that is spatio-temporally masked with a known mask $\mathbf{m}$:
$$\min_{\Theta} \mathcal{L}\bigg((\mathbf{m}\odot f_\Theta(\mathbf{z})), \mathbf{m}\odot\hat{\mathbf{x}}\bigg)$$Similar to the denoising experiment, we used the $\ell_2$ loss for $\mathcal{L}$ and performed the optimization for 2000 iterations. We construct masked observations by randomly introducing zero masks of duration varying between $0.1s$ to $0.25s$, such that all frequency components within the mask are zeroed out.

\noindent\textbf{Metrics}. For evaluation, we used the Spectral-SNR~\cite{spiertz2009source, virtanen2003sound}, a measure of the quality of spectrogram reconstruction, and the RMS Envelope distance~\cite{morgado2018self}. 

\noindent\textbf{Findings}. Table \ref{table:inp} compares the proposed approach against the baselines, using results from 50 examples in each dataset. It can be observed that, our approach consistently outperforms existing methods ($2.5$dB improvement in SNR on average) and introduces statistically meaningful patterns into the masked regions. This clearly demonstrates the efficacy of the proposed audio prior. 

\subsection{Source Separation}
We address the task of two source separation by adopting a formulation similar to~\cite{gandelsman2019double} - We use two audio priors aim to reconstruct the constituent sources $\{\hat{\mathbf{s}}_1, \hat{\mathbf{s}}_2\}$ and another prior to synthesize a mask $\mathbf{m}$, that can be used to mix the constituent sources to create the mixture audio, $\mathbf{m} \hat{\mathbf{s}}_1 + (1-\mathbf{m})\hat{\mathbf{s}}_2$. Similar to other previous restoration tasks, we use only a single observation (underdetermined setting) to estimate the sources. We synthesized 50 mixtures by randomly sampling and combining audio clips from the drums and the piano datasets and adopted loss functions from~\cite{tian2019deep}.  

\noindent{\textbf{Metrics}}.  We used the signal-to-distortion ratio (SDR), signal-to-interference ratio (SIR)~\cite{SiSEC18}, Spectral SNR and the RMS envelope distance metrics for evaluation. 

\noindent{\textbf{Findings}}. Table \ref{table:source} illustrates the performance of the DAP design choices. We find that, without increasing the complexity of the audio prior construction, the performance of U-Net architectures can be significantly improved through the proposed strategies. From our results, the effectiveness of the proposed audio prior design even with challenging inverse problems is clearly evident.

\section{Conclusion}
In this paper, we proposed a new deep audio prior construction that employs a carefully engineered convolutional architecture to produce significant performance gains in unsupervised audio restoration problems. In particular, we found that audio priors can be vastly improved through dilated convolutions with an exponential dilation schedule and dense connections. While the former strategy effectively increased the receptive fields for feature extraction, the latter supported a more principled learning of multi-scale features. As demonstrated by our experiments a suite of ill-posed audio restoration problems, the proposed approach provided meaningful signal priors to regularize this optimization process. 

\section{Acknowledgements}
This work was performed under the auspices of the U.S. Department of Energy by the Lawrence Livermore National Laboratory under Contract No. DE-AC52-07NA27344, Lawrence Livermore National Security, LLC. This document was prepared as an account of the work sponsored by an agency of the United States Government. Neither the United States Government nor Lawrence Livermore National Security, LLC, nor any of their employees makes any warranty, expressed or implied, or assumes any legal liability or responsibility for the accuracy, completeness, or usefulness of any information, apparatus, product, or process disclosed, or represents that its use would not infringe privately owned rights. Reference herein to any specific commercial product, process, or service by trade name, trademark, manufacturer, or otherwise does not necessarily constitute or imply its endorsement, recommendation, or favoring by the United States Government or Lawrence Livermore National Security, LLC. The views and opinions of the authors expressed herein do not necessarily state or reflect those of the United States Government or Lawrence Livermore National Security, LLC, and shall not be used for advertising or product endorsement purposes.

\bibliographystyle{IEEEtran}

\bibliography{mybib}


\end{document}